\documentclass[12pt]{article}
\usepackage{amssymb}
\usepackage{axodraw}
\usepackage{rotate}
\usepackage{epsfig}
\usepackage{wrapfig}
\usepackage{floatflt}

\begin{document}
\setcounter{page}{0}
\thispagestyle{empty}

\newcommand{\sanc}{{\tt SANC} }
\newcommand{\bqa}{\begin{eqnarray}}
\newcommand{\eqa}{\end{eqnarray}}
\newcommand{\Litwo}{\mbox{${\rm{Li}}_{2}$}}
\newcommand{\sss}[1]{\scriptscriptstyle{#1}}
\newcommand{\nll}{\nonumber\\}
\def\GF {G_{\sss F}}
\def\gw {\Gamma_{\sss W}}
\def\gz {\Gamma_{\sss Z}}
\def\mw {M_{\sss W}}
\def\mz {M_{\sss Z}}
\def\mh {M_{\sss H}}
\def\stw{s_{\sss W}}
\def\ctw{c_{\sss W}}
\newcommand{\GeV}{\unskip\,\mathrm{GeV}}
\newcommand{\MeV}{\unskip\,\mathrm{MeV}}
\newcommand{\hsm}{\hspace*{-1mm}}
\newcommand{\mpar}[1]{{\marginpar{\hbadness10000%
                      \sloppy\hfuzz10pt\boldmath\bf#1}}%
                      \typeout{marginpar: #1}\ignorespaces}
\newcommand{\tmW}{{\widetilde{M}}_{\sss W}}
\def\href#1#2{#2}

$\,$
\vspace*{-1cm}

\begin{flushright}
{\tt IFJPAN-V-2005-07}\\
{\tt hep-ph/0703043} 
\end{flushright}

\vspace*{\fill}

\begin{center}

{\LARGE\bf SANCnews: Sector 4$f$, Charged Current}

\vspace*{\fill}

{\bf A. Arbuzov$^{1,2}$, D. Bardin$^{2}$, S. Bondarenko$^{1,2}$,
P. Christova$^{2}$, \\
L.~Kalinovskaya$^{2}$, G. Nanava$^{3}$, R. Sadykov$^{2}$,
W. von Schlippe$^{4}$}

\vspace*{\fill}

{\normalsize{\it 
$^{1}$ Bogoliubov Laboratory of Theoretical Physics, 
JINR, Dubna, RU-141980,  Russia \\
$^{2}$ Dzhelepov Laboratory of Nuclear Problems,
JINR, Dubna, RU-141980,  Russia \\
$^{3}$ IFJ, PAN, Krakow, Poland \\
$^{4}$ PNPI, St. Petersburg, RU-188300, Russia}
\vspace*{13mm}
}
\end{center}

\begin{abstract}{In this paper we describe the implementation of the charged current decays of 
the kind $t\to b l^+\nu_l(\gamma)$ into framework of \sanc system. 
All calculations are done taking into account one-loop electroweak correction in the Standard 
Model. The emphasis of this paper is done on the presentation of numerical results. 
Various distributions are produced by means of a Monte Carlo integrator and event generator. 
Comparison with the results of CompHEP and PYTHIA packages are presented for the Born and hard 
photon contributions. The validity of the cascade approximation at one-loop level is also studied.}
\end{abstract}

\vspace*{\fill}

\centerline{\em Submitted to EPJC}

\vspace*{\fill}

\footnoterule
\noindent
{\footnotesize \noindent
This work is partly supported by INTAS grant $N^{o}$ 03-51-4007 \\ 
and by the EU grant mTkd-CT-2004-510126 in partnership with the
CERN Physics Department and by the Polish Ministry of Scientific Research and
Information Technology grant No 620/E-77/6.PRUE/DIE 188/2005-2008.}

\section{Introduction}
In this paper we describe a further application of the computer system \sanc {\em Support of 
Analytic and Numerical calculations for experiments at Colliders} intended for semi-automatic 
calculations of realistic and pseudo-observables for various processes of elementary particle 
interactions at the one-loop precision level
(see Ref.\cite{Andonov:2004hi} and references therein).

 Here we concentrate on the implementation of the 4 legs decay $t\to b + l^{+} + \nu_l$ as a 
typical example of the charged current (CC) decay of the kind $F\to f+f_1+\bar{f^{'}_1}$ where 
$F$ and $f$ stand for massive fermions and $f_1$ and $\bar{f^{'}_1}$ for massless fermions. 
In this paper we continue to present the 
physical applications of the \sanc system, started in Ref.\cite{Arbuzov:2005dd} rather than 
present extension of the system itself as continued in Ref.\cite{Bardin:2005dp}.

 According to the SM the dominant channel of top quark decay is $t\to b W^+$ with a branching 
ratio of $99.9\%$. The decay branching ratio of the $W$ boson into leptons is
Br$(W \to l^+\nu_l)\approx 11 \%$ Ref.\cite{item:ATLAS}.
Therefore, the semileptonic decays $t\to b l^+\nu_l$ ($l^+ \equiv e^+,\,\mu^+,\,\tau^+$)
amount to approximately $1/3$ of all top quark decays.

 This paper is devoted to the complete one-loop QED and EW radiative corrections (EWRC) to the 
4 legs semileptonic top quark decay $t\to b l^+\nu_l (\gamma)$. 
The calculation of QCD corrections in \sanc
for these 3 and 4 legs top decays is presented in Ref.\cite{Andonov:2006un}.

 EW and QCD radiative corrections to the 3 legs decay $t\to bW^+$ were first calculated
in the papers\cite{Denner:1990ns,Eilam:1991iz,Irwin:1990ka} and relevant issues may be found in 
Refs.\cite{Kuruma:1992if,Lampe:1995xb,Oliveira:2001vw,Fischer:2001gp,Do:2002ky,Smith:1994id,Mrenna:1991wd};
even two-loop QCD corrections are known \cite{Chetyrkin:1999br,item_sl,Cao:2004yy}.
However, we are not aware of papers where the 4 legs top decay $t\to b l^{+}\nu_l$ would be considered 
at one-loop.

 The results for the Born level decay width, presented in this paper, are compared with the calculation 
performed by means of CompHEP~\cite{Boos:2004kh} and PYTHIA~\cite{Sjostrand:2006za} packages and  
those for the 5 legs accompanying bremsstrah\-lung --- with the results of CompHEP. 
 We also discuss briefly how our results for the one-loop EW corrections are compared with results 
existing in the literature. The validity of the cascade approximation is also studied.

 The paper is organized as follows.
In section~\ref{Calc} we briefly recall the calculational scheme adopted in \sanc. 
The Born level is given in section~\ref{Born} and the one-loop EW corrections in 
section~\ref{Oneloop}. Various numerical results are collected in section~\ref{Numerics}.
In section~\ref{Cascade} we discuss the cascade approach to the problem, and in 
section~\ref{Concl} we present some conclusions.
 We assume that the reader may run \sanc as described in section 6 of Ref.\cite{Andonov:2004hi}
in order to see all relevant formulae which are not presented in this paper and get the corresponding
numbers.

\section{Calculation scheme\label{Calc}}
Recall that \sanc performs calculations starting from the construction of EW form factors (FF) 
which parameterize the covariant amplitude (CA) and helicity amplitudes (HA) of a process. 
From HA, the {\tt s2n} software produces the FORTRAN codes for them and then
the differential decay width is computed numecally. These codes can be further used in MC generators 
and integrators. 
The amplitudes (CA and HA) for the 4 legs top and antitop decays are presented in Ref.\cite{Andonov:2004hi}.

\begin{figure}[!ht]
\begin{center}
\includegraphics[width=0.45\textwidth]{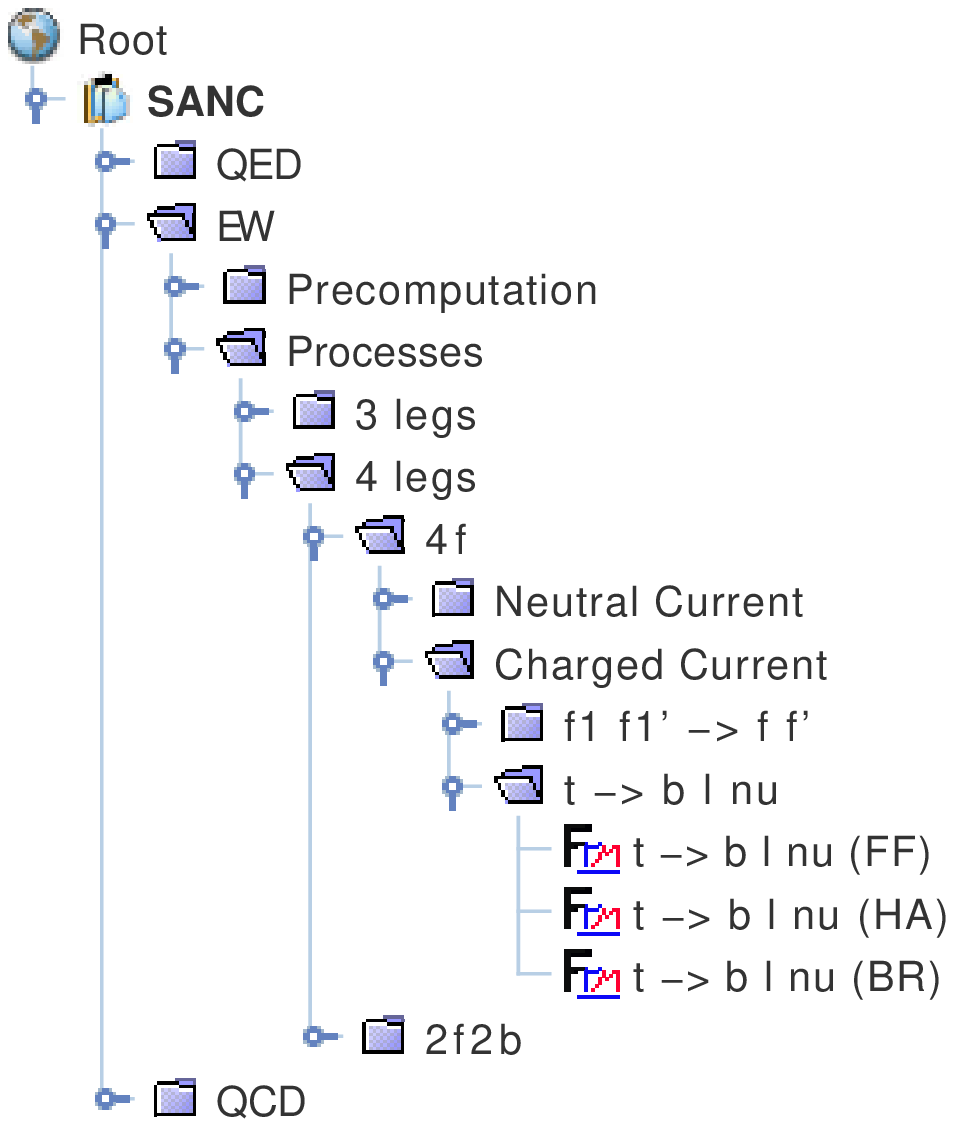}
\end{center}
\vspace*{-5mm}
\caption{\sanc tree for $t \to b l^+ \nu_l$ decay.}
\label{fig3}
\end{figure}

 These two ingredients, together with accompanying bremsstrahlung (BR)  are accessible via menu 
sequence {\bf SANC $\to$ EW $\to$ Processes $\to$ 4 legs $\to$ 4f $\to$ Charged current
$\to$ t $->$ b l nu $\to$ t $->$ b l nu (FF,~HA,~BR)}, see Fig.\ref{fig3}. 
A FORM (see Ref.\cite{Vermaseren:2000nd})
module, loaded at the end of this chain computes on-line the FF, HA and BR, respectively.
For more detail see section 2.5 of the \sanc
description in Ref.\cite{Andonov:2004hi} and the book\cite{item:book}.
\vspace*{-5mm}

The total one-loop width, $\Gamma^{\mathrm{1-loop}}$, of the decay $t\to b l^+\nu_l(\gamma)$ 
can be subdivided into the following terms:
\bqa
\Gamma^{\mathrm{1-loop}}&=&\Gamma^{\mathrm{Born}}+\Gamma^{\mathrm{virt}}(\lambda)
+\Gamma^{\mathrm{real}}(\lambda,\bar{\omega}),
\nll
\Gamma^{\mathrm{real}}(\lambda,\bar{\omega})&=&\Gamma^{\mathrm{soft}}(\lambda,\bar{\omega})
+\Gamma^{\mathrm{hard}}(\bar{\omega}).
\label{fourcont}
\eqa
Here $\Gamma^{\mathrm{Born}}$ is the decay width in the Born approximation, 
$\Gamma^{\mathrm{virt}}$ is the virtual contribution, $\Gamma^{\mathrm{soft}}$ and 
$\Gamma^{\mathrm{hard}}$
are the contributions due to the soft and hard photon emission respectively. 
The auxiliary parameter $\bar{\omega}$ separates the soft and hard photon contributions 
and the parameter $\lambda$ ("photon mass"), which enters the
virtual and soft contributions, regularizes the infrared divergences.

We present numbers, collected for the standard \sanc {\tt INPUT}, PDG(2006)~\cite{PDG2006}:
\bqa \nonumber
\begin{array}[b]{lcllcllcllcllcl}
G_{\sss F} & = & 1.16637\cdot 10^{-5}\GeV^{-2}, &\alpha(0) &=& 1/137.03599911 \\
\mw & = & 80.403\GeV, &
\gw & = & 2.141\GeV, \\
\mz & = & 91.1876\GeV,& 
\gz & = & 2.4952\GeV, \\
\mh & = & 120\GeV,    &\\
m_e & = & 0.51099892\cdot 10^{-3}\GeV,& m_u & = & 62\;\MeV & m_d & = & 83\;\MeV,\\
m_{\mu}&=&0.105658369\GeV,& m_c & = &1.5\;\GeV, & m_s & = & 215\;\MeV,\\
m_{\tau}&=&1.77699\GeV,& m_b & = & 4.7\;\GeV, & m_t & = & 174.2\;\GeV. \\
\end{array}
\label{sanc2006}
\eqa

The coupling constants can be set to different va\-lu\-es according to the different input 
parameter
schemes. They can be directly identified with the fine-structure constant $\alpha(0)$ together 
with $e/g=\stw$ and $\ctw=\mw/\mz$. This choice is called $\alpha$ scheme. 
Another one, the $G_F$ scheme, makes use of the Fermi constant and the quantity $\Delta r$. 
Note, we do not iterate the equation for $\Delta r$.
We use both schemes to produce numbers.

\section{Born-level process\label{Born}}
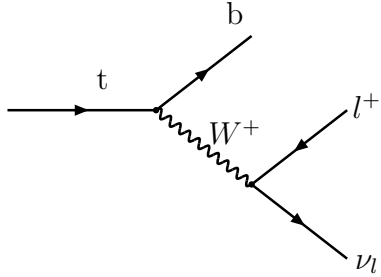
\begin{figure}[!h]
\begin{center}
\begin{picture}(200,125)(0,-25)
\SetScale{0.8}
\SetWidth{1.2}
\ArrowLine(5,85)(75,85)
\Vertex(75,85){1.5}
\ArrowLine(75,85)(120,120)
\Photon(75,85)(120,50){2}{10}
\Vertex(120,50){1.5}
\ArrowLine(165,85)(120,50)
\ArrowLine(120,50)(165,15)
\Text(40,80)[]{t}
\Text(90,60)[]{$W^+$}
\Text(90,105)[]{b}
\Text(140,70)[]{$l^+$}
\Text(140,10)[]{$\nu_l$}
\end{picture}
\vspace*{-20mm}
\end{center}
\caption{Feynman diagram for Born level process.\label{fig1}}
\end{figure}

In the Born approximation there is only one Feynman diagram for the decay $t\to b l^+\nu_l$
with one intermediate virtual $W^+$ boson, see Fig.\ref{fig1}.
\vspace*{1mm}

The differential decay rate reads:
\begin{equation}
d\Gamma^{\mathrm{Born}}=
\frac{1}{2m_t}{\overline{\sum_{spins}}}{|\mathcal{M}^{\mathrm{Born}}|^2}d\Phi^{(3)}\,,
\end{equation}

\noindent
where $\mathcal{M}^{\mathrm{Born}}$ is the amplitude of the process and $d\Phi^{(3)}$ 
is the differential three-body phase space:
\vspace*{-5mm}

\begin{equation}
d\Phi^{(3)}=\Phi^{(2)}_{1}\,d\Phi^{(2)}_{2}\,\frac{ds}{2\pi}\,,
\end{equation}
expressed in terms of the two-body phase spaces: 
\bqa
\Phi^{(2)}_{1}&=&\frac{1}{8\pi}\frac{\sqrt{\lambda(m_t^2,m_b^2,s)}}{m_t^2}\,,
\nll
d\Phi^{(2)}_{2}&=&\frac{1}{8\pi}\frac{\sqrt{\lambda(s,m_l^2,0)}}{s}\frac{1}{2}d\cos{\theta}\,.
\eqa
One can express the values $|\mathcal{M}^{\mathrm{Born}}|^2$ and $d\Phi^{(3)}$
via two independent variables: $s = -(p_l+p_{\nu})^2$ and $\cos{\theta}$, where $\theta$ is the
angle between $\vec{p_l}$ and $\vec{p_b}$ in the rest frame of the compound ($l^+$, $\nu_l$).
The limits of variation are:
\begin{equation}
{m_l^2}\leq s\leq {(m_t-m_b)^2}\,,\qquad -1\leq\cos{\theta}\leq +1\,.
\end{equation}
If the lepton mass is not ignored, then the $s$ and $\vartheta$ dependence of 
the Mandelstam variables $t$ and $u$ is given by 
\bqa
(t,u) &=& m_b^2+m_l^2+\frac{1}{2s}
    \Bigl[\left( s + m_l^2 \right) \left( m_t^2 - m_b^2 -s \right)
       \mp\left( s - m_l^2 \right) \sqrt{\lambda_s}\cos\theta\Bigr],
\eqa
where ${\lambda_s}={(m_t^2+m_b^2-s)^2-4 m_b^2 m_t^2}.$
\vspace*{-5mm}
\clearpage

\begin{figure}[!h]
\includegraphics[width=0.5\textwidth]{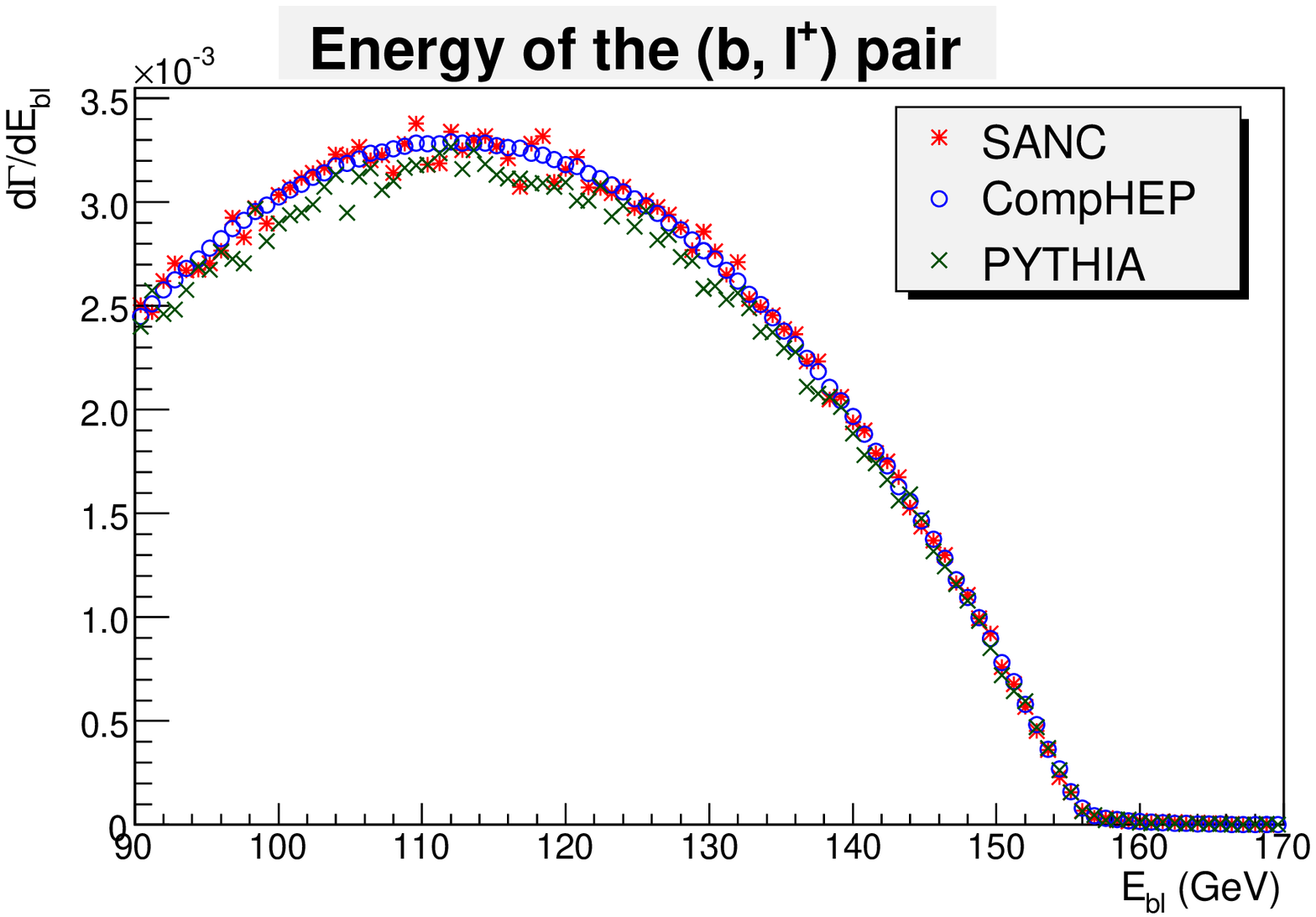}
\includegraphics[width=0.5\textwidth]{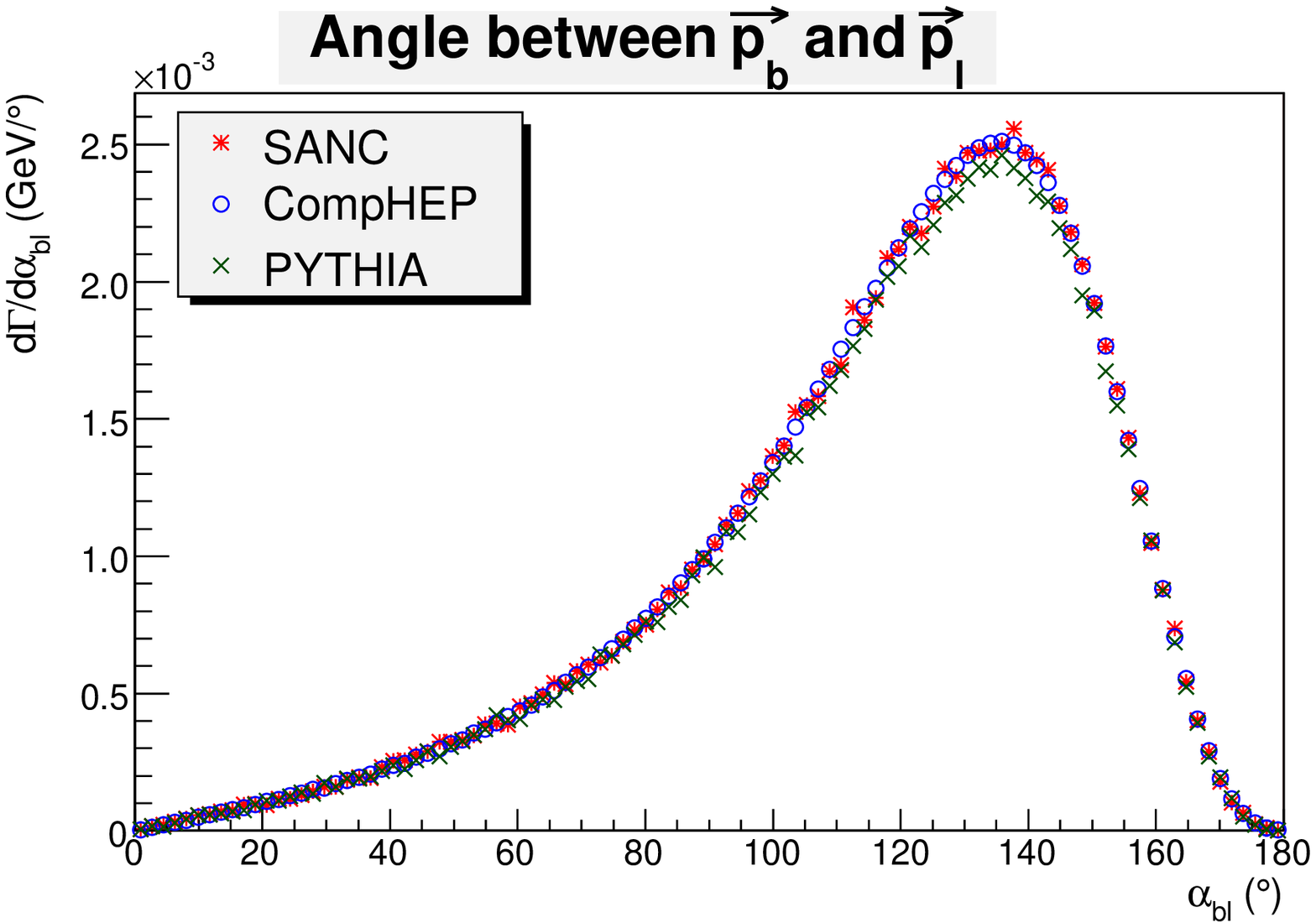}
\caption{Differential distributions for process $t\to b\mu^+\nu_\mu$
of the $b\mu^{+}$ pair energy and the angle between $\vec{p}_b$ and $\vec{p}_l$
produced with help of \sanc, CompHEP and PYTHIA.\label{fig2}}
\end{figure}

\begin{figure}[!t]
\includegraphics[width=0.5\textwidth]{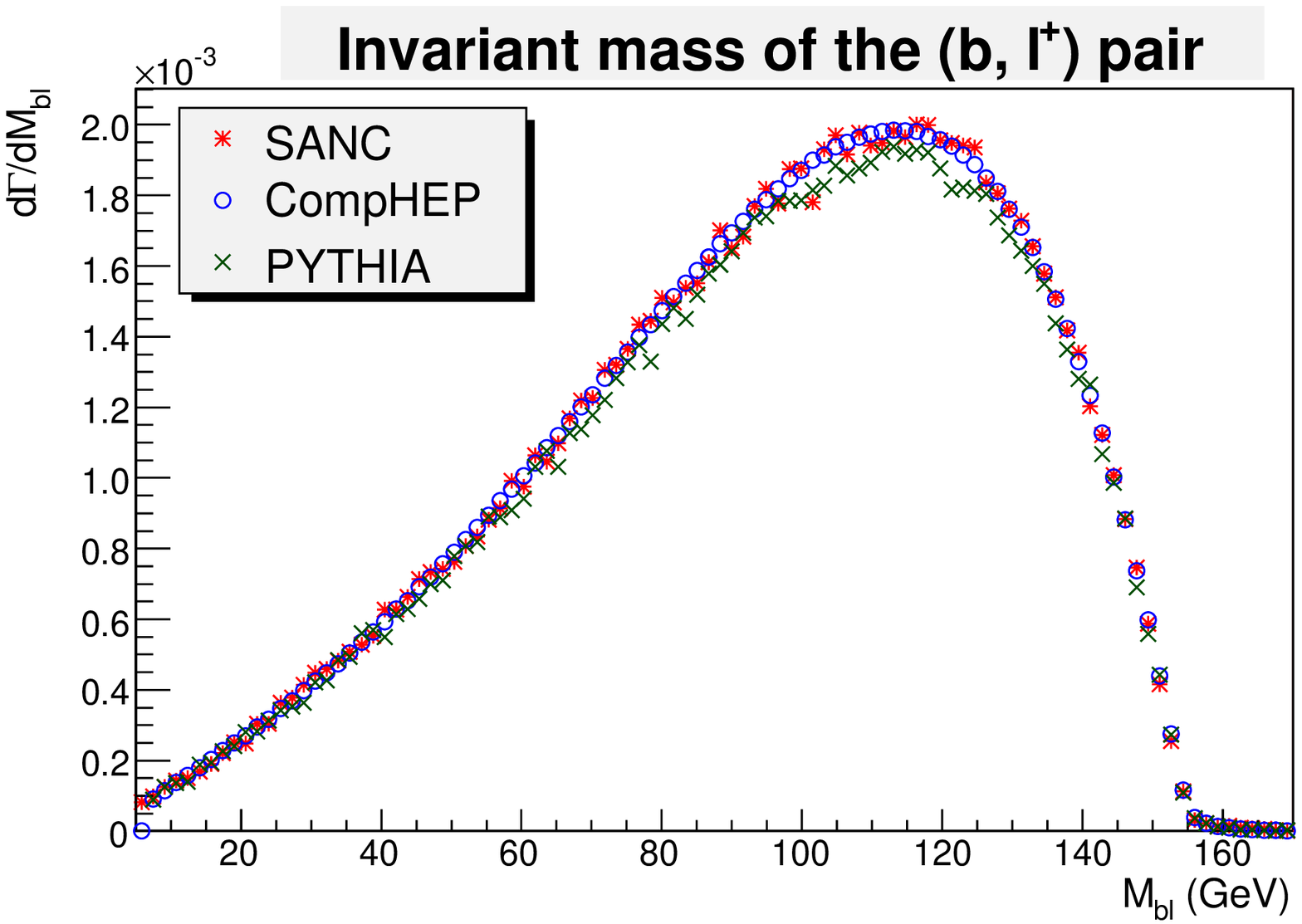}
\includegraphics[width=0.5\textwidth]{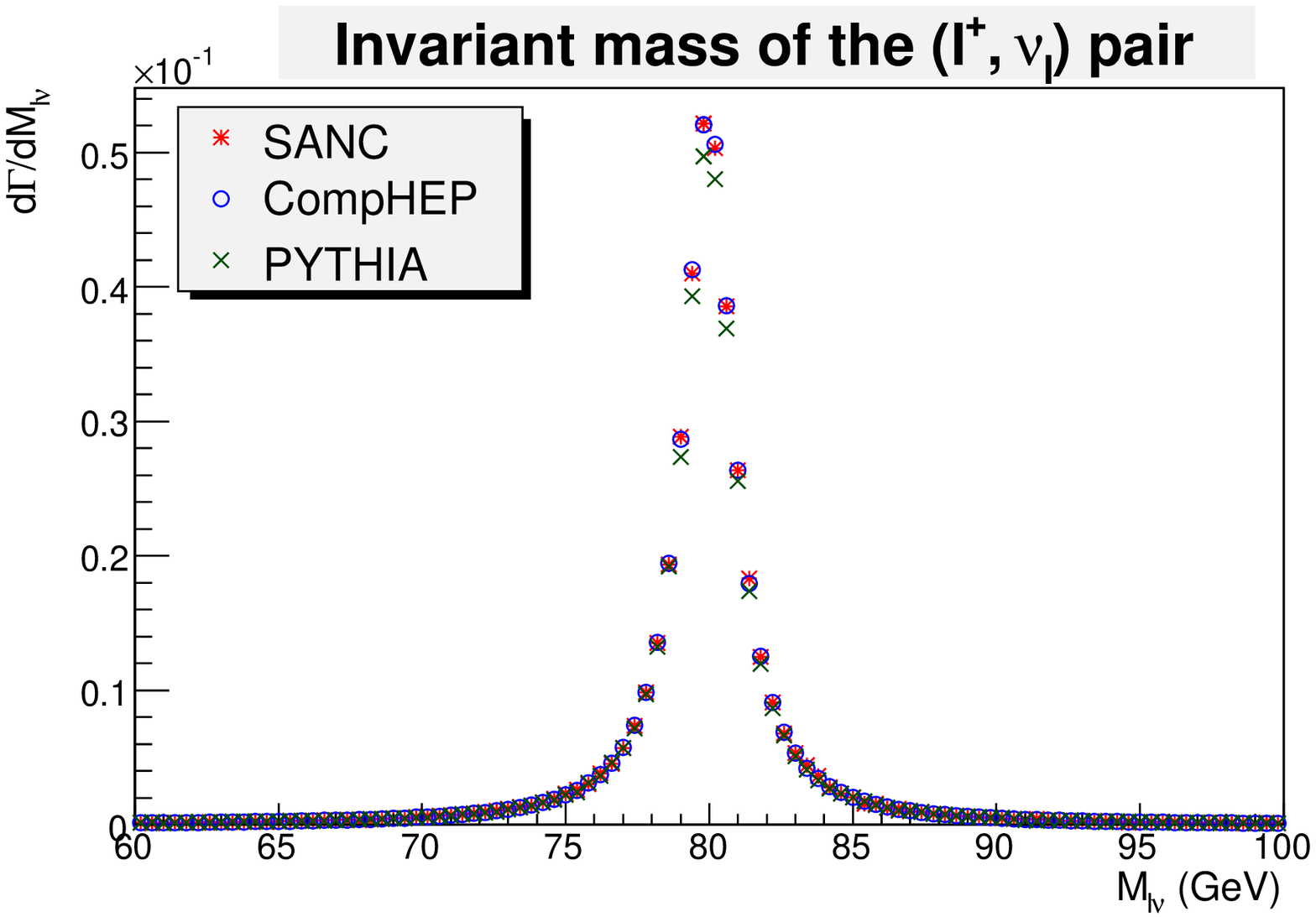}
\caption{Differential distributions for process $t\to b\mu^+\nu_\mu$
of invariant masses of $b\mu^{+}$ and $\mu^{+}\nu_{\mu}$ pairs
produced by \sanc, CompHEP and PYTHIA.\label{fig2p}}
\end{figure}

The result of the two-fold Monte Carlo integration is
shown in Table~\ref{tab1}. This calculation is performed by means of a
Monte Carlo integration routine based on the VEGAS algorithm Ref.\cite{Lepage:1977sw}. 
The numbers produced with help of Comp\-HEP and PYTHIA packages are also presented in the table.
\begin{table}[ht]
\begin{center}
\begin{tabular}{|c|c|c|}
\hline
\multicolumn{3}{|c|}{$\Gamma^{\mathrm{Born}}$,~GeV}\\
\hline
  \sanc &  CompHEP &  PYTHIA \\
\hline
$0.16936(1)$&$0.16935(1)$&$0.16782(1)$\\
\hline
\end{tabular}
\end{center}
\caption{Born-level decay width for decay $t \to b \mu^+ \nu_\mu$
produced by \sanc, CompHEP and PYTHIA.\label{tab1}}
\vspace*{-2mm}
\end{table}
The results of \sanc and CompHEP are in a good agreement, the deviation from PYTHIA appears due to
the difference in the defi\-nition of EW constants.
In addition to integration we use a Monte Carlo generator of unweighted events to produce
differential distributions. In Fig.\ref{fig2} we present some of these distributions and a comparison
with distributions, obtained with help of CompHEP and PYTHIA packages. We note, that the input 
parameters for this comparison were tuned to CompHEP.
In Figures~\ref{fig2}--\ref{fig2p} we show a triple comparison for the four distributions over 
various kinematical variables at the Born level.  
\vspace*{2mm}

 The figures demonstrate a very good agreement between \sanc and CompHEP 
and a fair agreement with PYTHIA.

\section{Radiative corrections\label{Oneloop}}
Radiative corrections can be subdivided into two parts: 
{\em virtual (one-loop)} corrections and {\em real (single photon emission)}.
The latter, in turn, is subdivided into {\em soft} and {\em hard} photon emission, see 
Eq.~(\ref{fourcont}).

\subsection{Virtual corrections} 
\begin{figure}[!h]
\centering
\includegraphics[width=0.8\textwidth]{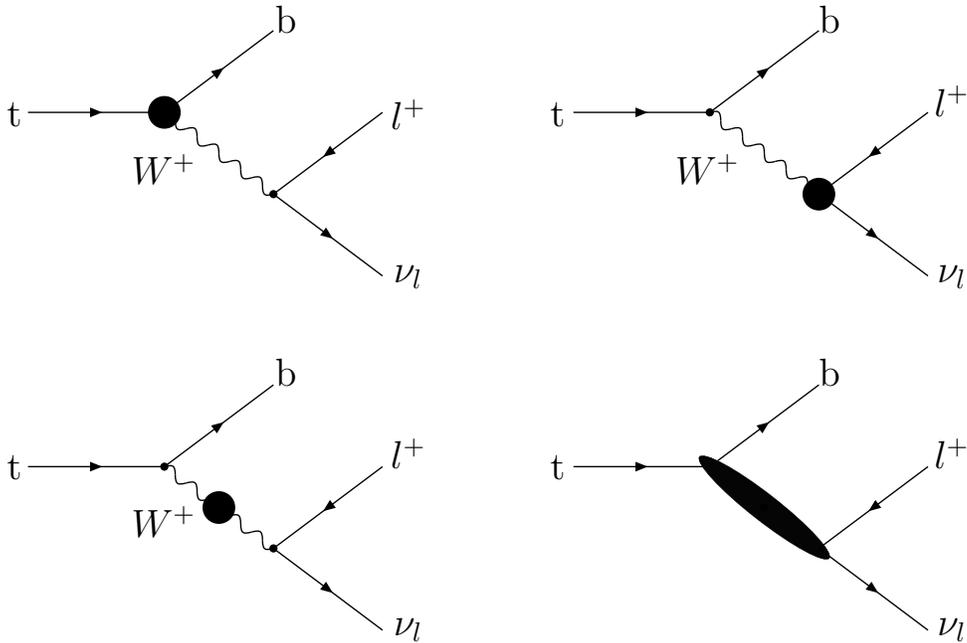}
\caption{Feynman diagrams for one-loop level decay.}
\label{fig4}
\vspace*{-17mm}
\end{figure}

Virtual corrections can be schematically represented by building block diagrams:
dressed vertices, self-ener\-gies and boxes, see~Fig.\ref{fig4}. They all, except the boxes,
include relevant counterterm contributions in the same spirit as described for the neutral
current (NC) case in Ref.\cite{Andonov:2002xc}. We also apply the recipe of 
Ref.\cite{Wackeroth:1996hz} to regularize the so-called ``on-mass-shell'' singularities.

 The virtual contribution is parameterized by scalar form factors which can be found 
in the ``{\tt SANC Output window}'' after a run of the FF-module on the top decay branch 
of the \sanc tree, see~Fig.\ref{fig3}.
\vspace*{-3mm}

\clearpage

\subsection{Real corrections}
The soft contribution is proportional to the Born level decay rate
and has the same phase space. Its explicit expression can also be found
in the ``{\tt SANC Output window}'' after \sanc-run of BR-module, see~Fig.\ref{fig3}.

\begin{figure}[!h]
\centering
\includegraphics[width=0.8\textwidth]{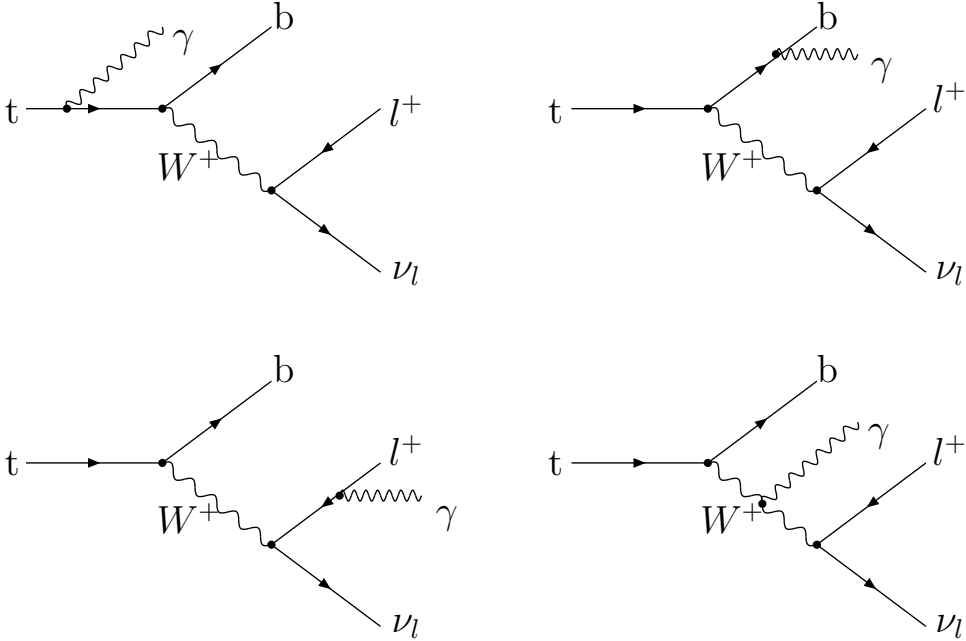}
\caption{Feynman diagrams for hard photon emission.}
\label{fig6}
\end{figure}

For hard photon emission there are four tree-level Feynman diagrams (see Fig.\ref{fig6}).
One dia\-gram corresponds to emission from the initial state, two diagrams describe the final
state radiation and the remaining diag\-ram corresponds to radiation from the intermediate
$W^+$ boson.

 Hard bremsstrahlung in $t(p_1)\to b(p_2) + l^{+}(p_3)+\nu_l(p_4)+\gamma(p_5)$ 
has the four-body phase space:
\bqa
d\Phi^{(4)}= \Phi^{(2)}_1  d\Phi^{(2)}_2  d\Phi^{(2)}_3 \frac{ds_{25}}{2\pi}\frac{ds_{34}}{2\pi}\,,
\eqa
where the three two-body phase spaces are:
\bqa
\Phi^{(2)}_1&=&\frac{1}{8\pi}\frac{\sqrt{\lambda(m_t^2,s_{25},s_{34})}}{m_t^2}\,,
\nll
d\Phi^{(2)}_2&=&\frac{1}{8\pi}\frac{\sqrt{\lambda(s_{25},m_b^2,0)}}{s_{25}}
\frac{1}{2}d\cos{\theta_1}\,,
\nll
d\Phi^{(2)}_3&=&\frac{1}{8\pi}\frac{\sqrt{\lambda(s_{34},m_l^2,0)}}{s_{34}}
\frac{1}{2}d\cos{\theta_2}d\phi_2\,.
\eqa

\begin{figure}[!ht]
\centering
\includegraphics[width=0.8\textwidth]{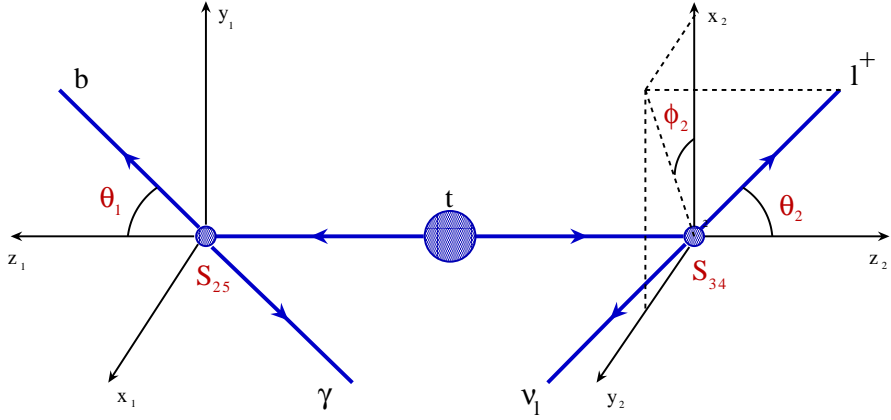}
\caption{Kinematical diagram for hard photon emission.}
\label{fig5}
\end{figure}

The kinematics and meaning of variables are illustrated in Fig.\ref{fig5}.

The total decay rate for the hard process is represented by a 5-fold integral over
$s_{25},\,s_{34},\,\cos{\theta_1},$ $\cos{\theta_2},$ $\phi_2$,
varying within the following limits:
\begin{eqnarray}
&m^2_b\leq s_{25}\leq (m_t-m_l)^2,
\nll
&m^2_l\leq s_{34}\leq (m_t-\sqrt{s_{25}})^2,
\nll
&-1\leq \cos{\theta_1}\leq +1\,,
\nll
&-1\leq \cos{\theta_2}\leq +1\,,
\nll
& 0\leq \phi_2\leq 2\pi\,.
\end{eqnarray}
The matrix element of Fig.\ref{fig6} and the kinematics described in this section
are basis for the \sanc Monte Carlo generator.

\section{Numerical results\label{Numerics}}
\subsection{Comparison of hard bremsstrahlung between \sanc and Comp\-HEP.}
We begin by presenting the results of the Monte Carlo integration 
of the hard photon contributions derived with the help of \sanc and CompHEP 
as presented in Table~\ref{tab2}.
\begin{table}[ht]
\begin{center}
\begin{tabular}{|c|c|c|}
\hline
$\bar\omega$, GeV&
$\Gamma^{\mathrm{hard}}$, $10^{-2}$GeV&
$\Gamma^{\mathrm{hard}}$, $10^{-2}$GeV\\
& CompHEP & \sanc \\
\hline
10       &0.2578(2)&0.2592(2)\\
1        &0.6982(3)&0.8582(2)\\
$10^{-1}$&0.8538(3)&1.5000(3)\\
$10^{-2}$&0.9628(4)&2.1495(3)\\
$10^{-3}$&1.0730(4)&2.8005(4)\\
$10^{-4}$&1.1809(3)&3.4525(4)\\
\hline
\end{tabular}
\end{center}
\caption{Comparison for hard emission produced by \sanc and CompHEP systems for 
$E_\gamma\geq\bar{\omega}$.}
\label{tab2}
\end{table}

There is a significant difference between two sets of numbers and this difference
increases with decreasing  $\bar{\omega}$. This difference is due to the approximate
representation of the W boson propagators implemented in CompHEP;
in CompHEP the complex propagator is used in a real 
representation:\footnote{Here we use exceptionally the metric $p^2=M^2$.}
\begin{equation}
\frac{1}{p^2-\mw^2 + i\mw\gw} \to \frac{p^2-\mw^2}{(p^2-\mw^2)^2 + \mw^2\gw^2}\,.
\label{compprop}
\end{equation}

\begin{figure}[t!]
\includegraphics[width=0.5\textwidth]{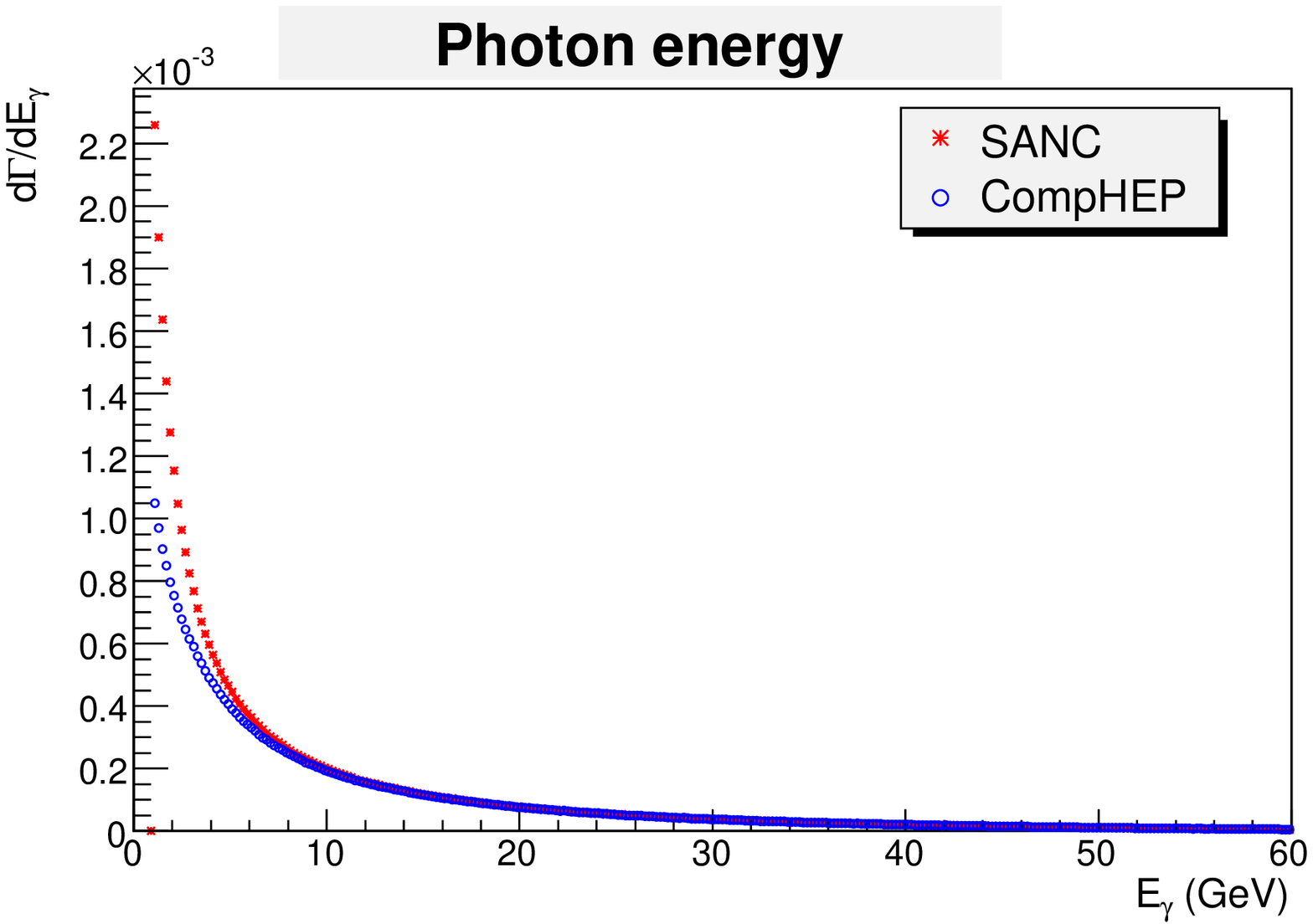}
\includegraphics[width=0.5\textwidth]{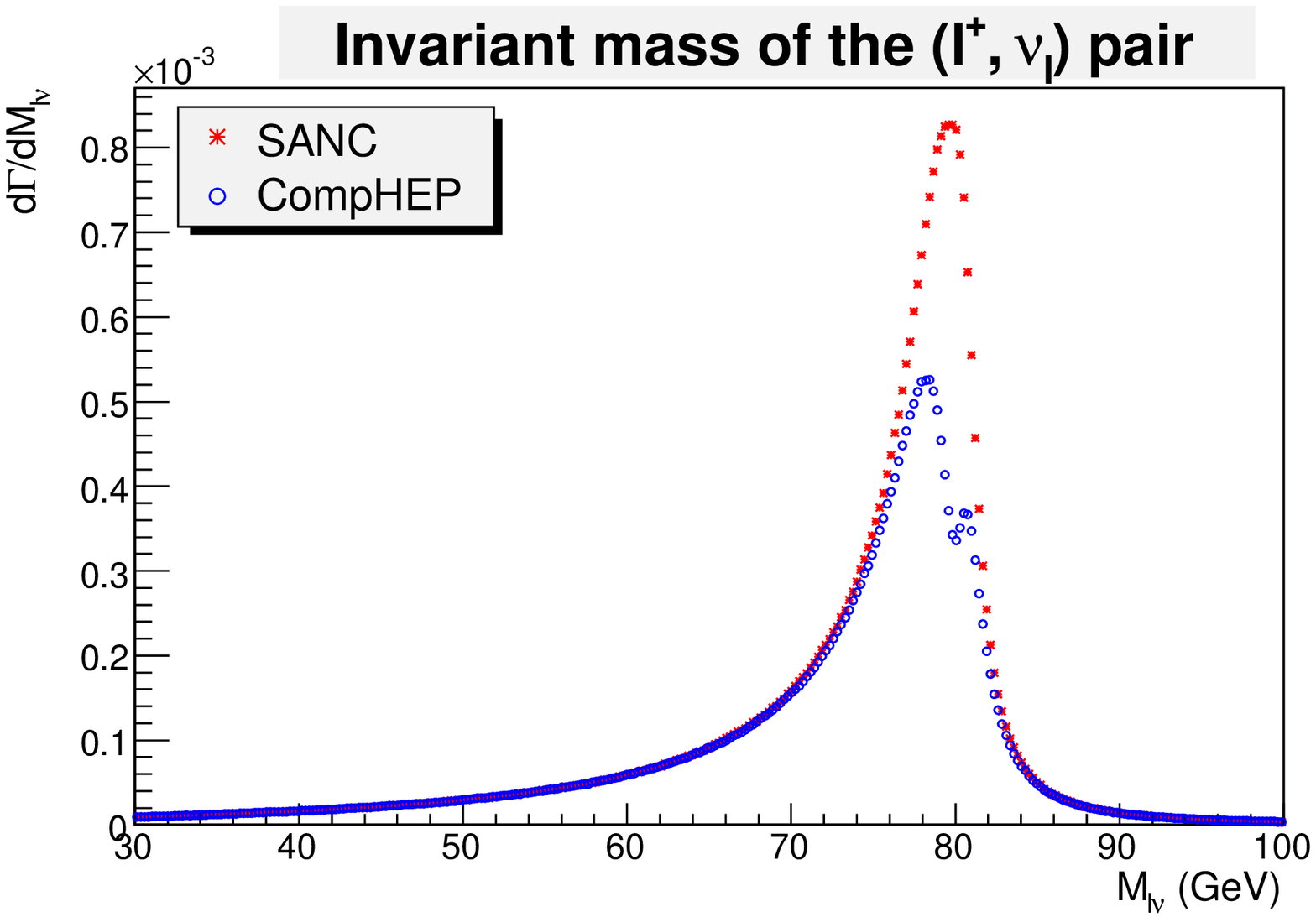}
\includegraphics[width=0.5\textwidth]{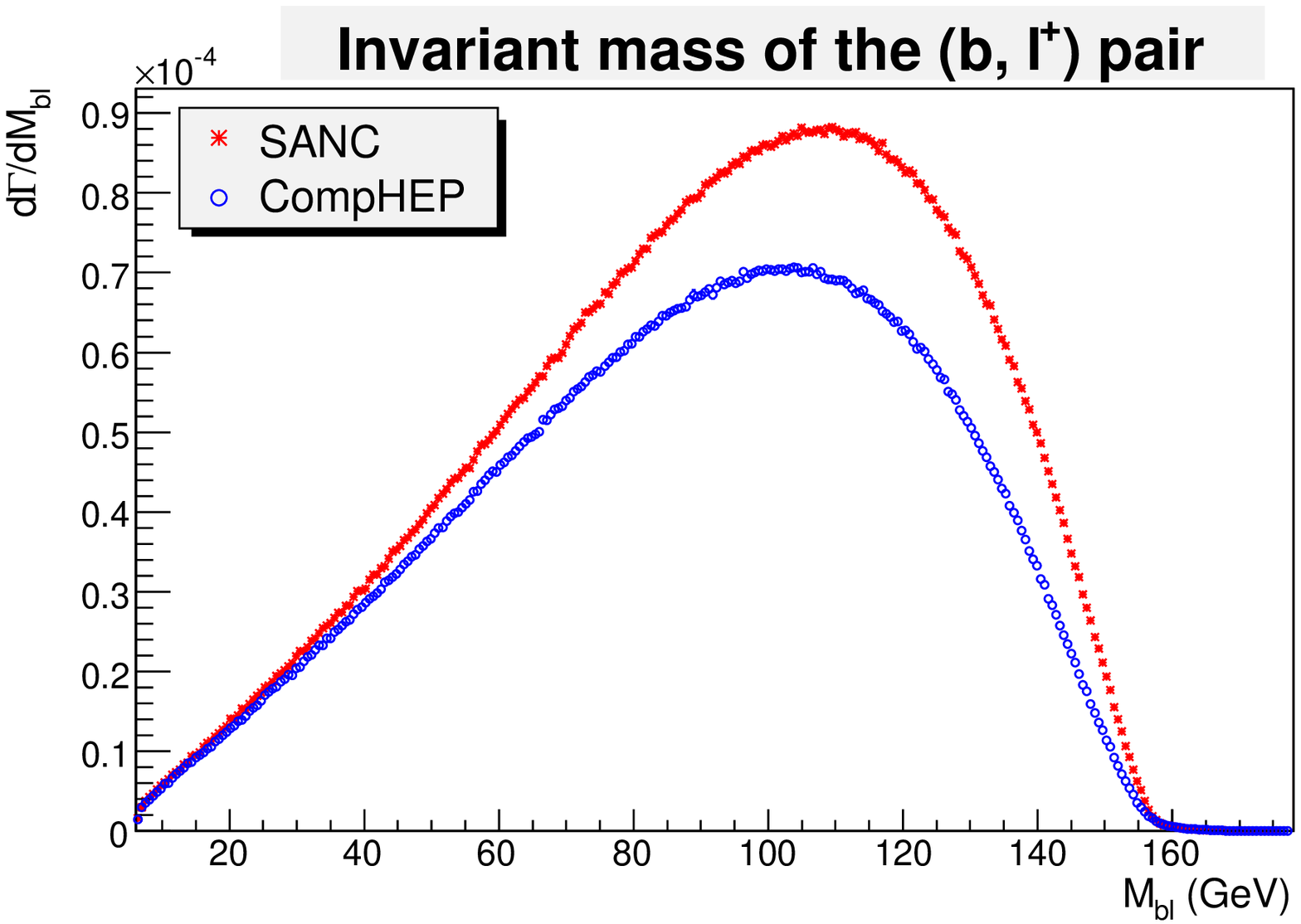}
\includegraphics[width=0.5\textwidth]{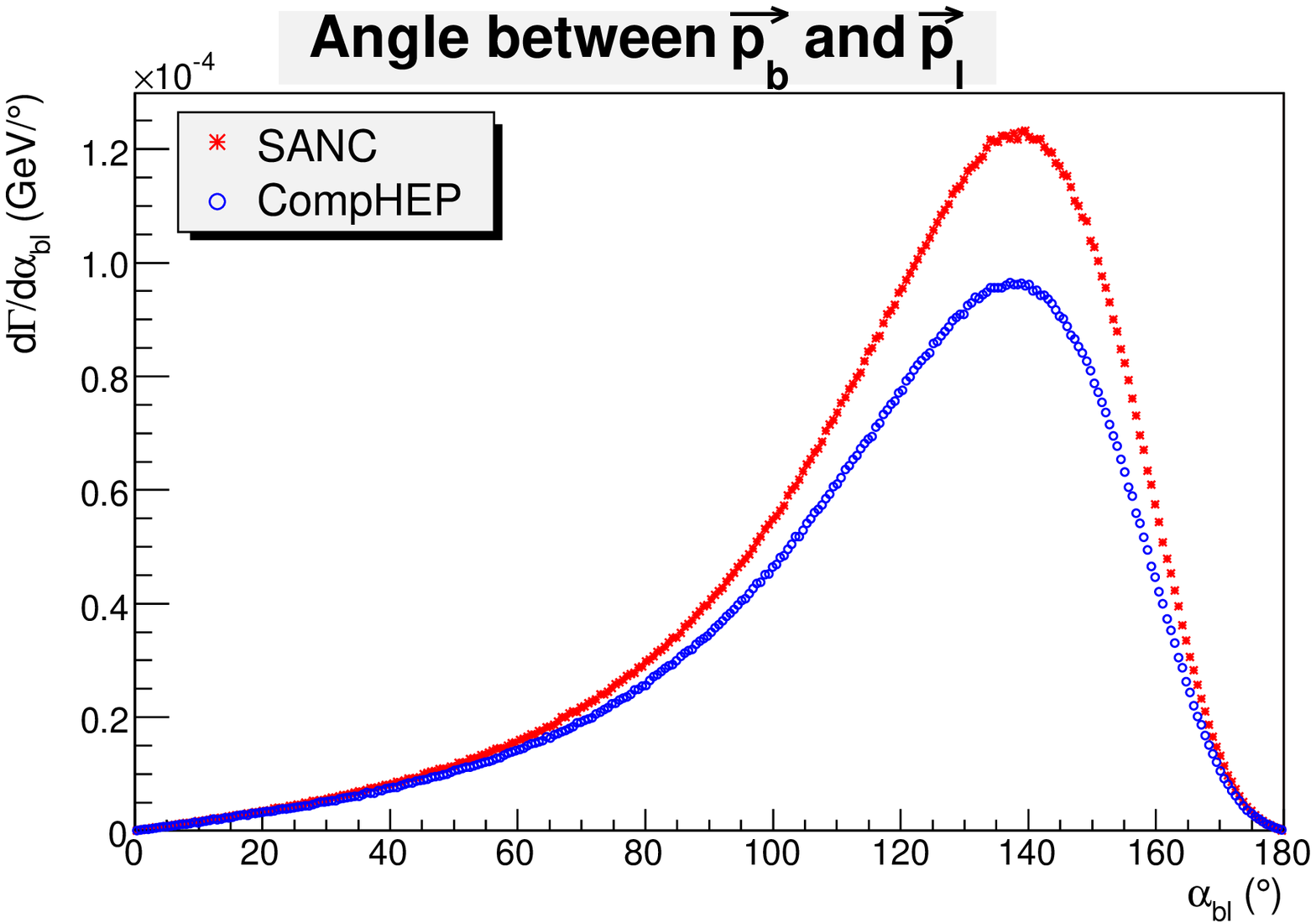}
\caption{Differential distributions for hard photon emission process $t\to b\mu^+\nu_l\gamma$
with $E_{\gamma}\geq 1$~GeV.}
\label{fig7}
\end{figure}

This assumption will not lead to a noticeable departure from the correct result with the exception 
of the case when we have the product of two different W propagators 
(i.e. with different virtualities).
In this case it is necessary to make a substitution that corrects this assumption:
\bqa
&&\frac{p_1^2-\mw^2}{(p_1^2-\mw^2)^2+\mw^2\gw^2}\,\,\frac{p_2^2-\mw^2}{(p_2^2-\mw^2)^2+\mw^2\gw^2}
\to
\nll
&&\frac{p_1^2-\mw^2}{(p_1^2-\mw^2)^2+\mw^2\gw^2}\,\,\frac{p_2^2-\mw^2}{(p_2^2-\mw^2)^2+\mw^2\gw^2}
\nll
&+&\frac{M_m^2\gw^2}{((p_1^2-\mw^2)^2+\mw^2\gw^2)((p_2^2-\mw^2)^2+\mw^2\gw^2)}\,.
\eqa
We can explicitly observe the difference in Fig.\ref{fig7}, where we present the various 
differential distributions. As indicated in the upper two pictures the difference is to be 
seen in the region of soft photon emission near the resonance.

Note, that if we use recipe~(\ref{compprop}) in \sanc, then we simulate the CompHEP distributions 
with a very good precision.

\subsection{Numerical results for the complete EWRC}
The results for the complete one-loop calculation of widths in $\alpha$ and $G_F$ schemes
and comparison with Born level widths are presented in the Tables~\ref{tab3} and 
\ref{tab4}.\footnote{NB: Although \sanc may produce all results exact in $b$-mass,
all numbers in this subsection and section~\ref{Cascade} are derived for $m_b\to 0$.}
 
\begin{table}[ht]
\begin{center}
\begin{tabular}{|l|l|l|l|}
\hline
l&
$\Gamma^{\mathrm{Born}}$, GeV&
$\Gamma^{\mathrm{1-loop}}$, GeV&
$\delta$, $\%$\\
\hline
$l^+$ &0.14948(1)& 0.16064(1)&7.47\\
\hline
\end{tabular}
\vspace*{2mm}
\caption{Born and one-loop decay width and percentage 
of the correction in $\alpha$ scheme.\label{tab3}}
\end{center}
\end{table}
\clearpage

\begin{table}[!h]
\begin{center}
\begin{tabular}{|l|l|l|l|}
\hline
l&
$\Gamma^{\mathrm{Born}}$, GeV&
$\Gamma^{\mathrm{1-loop}}$, GeV&
$\delta$, $\%$\\
\hline
$l^+$&0.16018(1)&0.16299(1)&1.75\\
\hline
\end{tabular}
\vspace*{2mm}
\caption{Born and one-loop decay width and percentage
of the correction in $G_F$ scheme.\label{tab4}}
\vspace*{-5mm}
\end{center}
\end{table}

There is practically no sensitivity to the lepton mass since it is neglected everywhere 
but in arguments of logs which, in turn, are vanishing due to the KLN theo\-rem.

\section{EWRC in cascade approximation\label{Cascade}}
It is interesting to compare results of the complete app\-roach with an approximate, ``cascade'' 
calculation based on the formula (we consider the case $l=e$):
\begin{eqnarray}
\Gamma_{t\to be\nu}=
\frac{\Gamma_{t\to Wb}\Gamma_{W\to e\nu}}{\gw}\,.
\label{cascad}
\end{eqnarray}
The input parameters are as in Eq.~(\ref{sanc2006}), except $m_b$ which is set to zero here,
and we present results in the $\alpha$ and $G_F$ schemes.
Consider first the validity of Eq.~(\ref{cascad}) at the Born level for a ({\it formal}) 
variation of $\gw$, see Table~\ref{tab5}:

\clearpage

\begin{table}[!t]
\begin{center}
\begin{tabular}{|c|c|c|c|c|}
\hline
&$\Gamma^{\mathrm{Born}}$, GeV    
&$\Gamma^{\mathrm{Born}}_{\mathrm{cascade}}$, GeV  
&$\delta, \%$ \\
\hline
 $\gw $    & 0.14948 & 0.15187 &  1.6 \\[-1mm]
 $\gw/10  $& 1.5163  & 1.5187  &  0.2 \\[-1mm]
 $\gw/10^2$& 15.185  & 15.187  &  0.01 \\[-1mm] 
 $\gw/10^3$& 151.87  & 151.87  &  0.00 \\
\hline
\end{tabular}
\end{center}
\caption{Comparison of Born widths without and with cascade approximation,
$\alpha(0)$-scheme.\label{tab5}}
\end{table}

The cascade approximation at the Born level improves rapidly with decreasing $\gw$.

Complete one-loop calculations are shown in Tables~\ref{tab6}--\ref{tab6gf}.

\begin{table}[!h]
\begin{center}
\begin{tabular}{|c|c|c|c|c|}
\hline
& $\Gamma^{\mathrm{Born}}$, GeV
& $\Gamma^{\mathrm{1-loop}}$, GeV
& $\delta, \%$ \\
\hline
 $\gw     $& 0.14949 & 0.16064 & 7.46  \\
\hline
\end{tabular}
\end{center}
\caption{Born and one-loop decay width and percentage
of the correction, $\alpha(0)$-scheme.\label{tab6}}
\end{table}

\begin{table}[!ht]
\begin{center}
\begin{tabular}{|c|c|c|c|c|}
\hline
& $\Gamma^{\mathrm{Born}}$, GeV
& $\Gamma^{\mathrm{1-loop}}$, GeV
& $\delta, \%$ \\
\hline
 $\gw     $& 0.16018 & 0.16299 & 1.75 \\
\hline
\end{tabular}
\end{center}
\caption{Born and one-loop decay width and percentage
of the correction, $G_F$-scheme.\label{tab6gf}}
\end{table}


Now turn to the one-loop version of cascade Eq.~(\ref{cascad}).
First, compute $\Gamma(t\to Wb)$ and $\Gamma(W\to e\nu)$ neglecting $\gw$ in all $W$ boson
propagators~(\ref{compprop}). So, in Eq.~(\ref{cascad}) the numerator does not depend on $\gw$.
In this ``naive'' variant of calculations it is sufficient to consider only one point over $\gw$, 
since the correction $\delta$ is a constant by construction.

\begin{table}[!h]
\begin{center}
\begin{tabular}{|c|c|c|c|c|c|}
\hline
&{$t\to Wb$}&{$W\to e\nu$}&{$t\to be\nu$ }\\[-1mm]
&           &             &{ cascade }    \\
\hline
$\Gamma^{\mathrm{Born}}$, GeV   & 1.4800 & 0.21970 & 0.15187 \\[-1mm]
$\Gamma^{\mathrm{1-loop}}$, GeV & 1.5466 & 0.22528 & 0.16274 \\[-1mm]   
$\delta, \%$                    & 4.49   & 2.54    & 7.15    \\    
\hline
\end{tabular}
\end{center}
\caption{Born, one-loop decay widths and percentage
of the correction in cascade approximation, $\alpha(0)$-scheme.\label{tab7}}
\end{table}

\begin{table}[!h]
\begin{center}
\begin{tabular}{|c|c|c|c|c|c|}
\hline
&{$t\to Wb$}&{$W\to e\nu$}&{$t\to be\nu$ }\\[-1mm]
&           &             &{ cascade }    \\
\hline
$\Gamma^{\mathrm{Born}}$,  GeV  & 1.5321 & 0.22742 & 0.16274 \\[-1mm]
$\Gamma^{\mathrm{1-loop}}$, GeV & 1.5572 & 0.22670 & 0.16488 \\[-1mm]   
$\delta, \%$                    & 1.64   & -0.32   & 1.31    \\    
\hline
\end{tabular}
\end{center}
\caption{Born, one-loop decay widths and percentage
of the correction in cascade approximation, $G_F$-scheme.\label{tab7gf}}
\end{table}
From Tables~\ref{tab6}--\ref{tab7gf} one sees that the complete and cascade one-loop 
calculations deviate conside\-rably. 
This hints to take into account effects of $\gw$ in cascade calculations more carefully.

At the end of this section we note that the percentage of EWRC correction for $t\to Wb$ decay
reasonably agrees with results given in Table~1 of Ref.\cite{Denner:1990ns},
even though we did not tune any parameters to achieve agreement.

\section{Conclusions \label{Concl}}
A study of the semileptonic top quark decay $t \to b l^+ \nu_l (\gamma)$ was presented.
We have computed the total one-loop electroweak corrections to this process with the aid of
the \sanc system.
Using a Monte Carlo integrator and an event generator that we have created
for this purpose, we specify the influence on
the decay width due to EWRC. These corrections are about $7.5 \%$ for
$\alpha$ scheme and approximately $1.8 \%$ for $G_F$ scheme. The comparison with the numbers of
CompHEP and PYTHIA packages was done at the tree level. During this comparison we found
noticeable deviation from the CompHEP package for soft photon emission in the resonance region.

We have studied the cascade approach to the problem under consideration.
We have shown that the ``naive'' approach with ``stable'' $W$'s is not precise enough. 
An improved treatment of the cascade approach with the complex $W$ mass will be presented elsewhere.
\vspace*{5mm}

\leftline{\Large\bf Acknowledgments}

\vspace*{2mm}
This work was partly supported by the INTAS grant 03-41-1007 (AA, DB, SB and LK)
by the RFBR grant 04-02-17192 (AA) 
and by the EU grant mTkd-CT-2004-510126 in partnership with the
CERN Physics Department and by the Polish Ministry of Scientific Research 
and Information Technology grant No 620/E-77/6.PRUE/DIE 188/2005-2008 (GN).
AA, DB, SB, LK and GN are indebted to the directorate of IFJ, 
Krakow, for hospitality which was extended to them in April--May 2005,
when an essential part of this study was done.

{}

\end{document}